# Schottky barrier lowering with the formation of crystalline Er silicide on *n*-Si upon thermal annealing

Nicolas Reckinger, Xiaohui Tang, and Vincent Bayot
*Microelectronics Laboratory, Université catholique de Louvain, Place du Levant 3, B-1348 Louvain-la-Neuve, Belgium*

Dmitri A. Yarekha, Sylvie Godey, Emmanuel Dubois, Xavier Wallart, and Guilhem Larrieu
*Institut d'Electronique, de Microélectronique et de Nanotechnologie, IEMN/ISEN UMR CNRS 8520, Avenue Poincaré, Cité Scientifique, 59652 Villeneuve d'Ascq Cedex, France*

Adam Łaszcz and Jacek Ratajczak
*Institute of Electron Technology, Al. Lotników 32/46, 02-668, Warsaw, Poland*

Pascal J. Jacques
*Laboratoire d'Ingénierie des Matériaux et des Procédés (IMAP), Université catholique de Louvain, Place Sainte Barbe 2, B-1348 Louvain-la-Neuve, Belgium*

Jean-Pierre Raskin
*Microwave Laboratory, Université catholique de Louvain, Place du Levant 3, B-1348 Louvain-la-Neuve, Belgium*

The evolution of the Schottky barrier height (SBH) of Er silicide contacts to *n*-Si is investigated as a function of the annealing temperature. The SBH is found to decrease substantially from 0.43 eV for as-deposited samples to reach its lowest value, 0.28 eV, at 450 °C. By x-ray diffraction, high resolution transmission electron microscopy and x-ray photoelectron spectroscopy, the decrease of the SBH is shown to be associated to the progressive formation of crystalline $ErSi_{2-x}$.

The rare-earth (RE) silicides are known to present the lowest Schottky barrier height (SBH) to *n*-Si ($\Phi_{Bn} \approx$ 0.28 eV) among all existing silicides[1,2]. This unique property makes them considerably attractive as source/drain (S/D) contacts for *n*-type Schottky barrier metal-oxide-semiconductor field-effect transistors. In the Schottky S/D technology, silicides are used as S/D materials instead of the traditionally implanted S/D, due to improved scalability and simpler fabrication techniques[3,4]. They are usually grown by solid-state reaction completed by thermal annealing after deposition. SBH data found in the literature for RE silicides are systematically given for the optimal temperature of formation for the hexagonal phase ($RESi_{2-x}$). However, no research group has ever reported how $\Phi_{Bn}$ evolves with the annealing temperature below that optimal temperature.

Because REs quickly oxidize[5,6] owing to a high affinity to oxygen, their silicides must be grown in special conditions, either *in situ* in ultrahigh vacuum[1,2] (UHV) or *ex situ* with a protective capping layer[7,8,9,10,11]. The possibility to obtain a low $\Phi_{Bn}$ of 0.28 eV with a Ti cap was demonstrated very recently.

In the present letter, we follow the evolution of the SBH of RE of Er silicide contacts grown by rapid thermal annealing (RTA) on *n*-Si with a Ti cap. We notice that $\Phi_{Bn}$ drastically decreases from 0.43 eV for the as-deposited sample to reach a minimal value of 0.28 eV at an optimal annealing temperature ($T_A$) of 450 °C. Relying upon experimental observations, we also provide a plausible reason for that dependence.

The initial bulk substrates are *n*-type lowly doped Si(100) wafers (concentration $\sim 10^{15}$ cm$^{-3}$). They are cleaned in sulfuric peroxide mixture. After rinsing in de-ionized water, they are dipped into 1% hydrofluoric (HF) acid to remove the grown oxide, rinsed, and dried with N$_2$. A mechanical mask fixed on the wafers is used to pattern face-to-face Schottky diodes separated by a Si series resistance $R_{Si}$ [see inset to Fig. 1(a)]. The wafers are immediately inserted into the evaporation chamber, to limit oxide re-growth in ambient air. The deposition is performed in an e-beam evaporator operating under UHV (base pressure $\sim 5 \times 10^{-9}$ mbar). 25 nm of Er and 10 nm of Ti are successively deposited without breaking the vacuum. The wafers are then brought to ambient conditions, transferred to a RTA system and thermally activated in forming gas (95% N$_2$ + 5% H$_2$) for 2 min, from 300 to 600 °C by steps of 50 °C. Face-to-face diodes are also left unannealed.

The SBH is determined from *I-V* curves of two face-to-face Schottky diodes measured by four-point contact technique at temperatures ranging from 290 down to 150 K with a step of 20 K. Depending on the measurement temperature range, *I* is either limited by $R_{Si}$ (high temperature) or by the Schottky contact (low temperature). Hence, the model of the face-to-face structure[12] used to fit the data (converted into Arrhenius plots) is adjusted by tuning $R_{Si}$ and $\Phi_{Bn}$ in the ohmic and Schottky regimes, respectively. The transport through the Schottky contact takes into account thermionic-field emission and barrier lowering due to image charge.

Figure 1(a) displays the dependence of the SBH on $T_A$. For the as-deposited sample, the extracted SBH amounts to 0.43 eV. After annealing at 300 °C, the SBH drops to 0.37 eV and is again substantially reduced to reach 0.28 eV upon annealing at 450 °C. Beyond that value, the SBH slightly increases what was attributed to the degradation of Er silicide owing to enhanced diffusion of oxygen under excessive $T_A$. The remainder of the text is devoted to the explanation of the SBH dependence in the low $T_A$ region (below 450 °C). To emphasize further the strong impact of $T_A$ on the electrical characteristics of the grown Er silicide, we illustrate in Fig. 1(b) the evolution of the experimental Arrhenius plot for the as-deposited, the 300 °C, and the 450 °C samples, respectively. We can observe that the transition temperature between the ohmic regime (positive slope in the Arrhenius plot) and the Schottky regime (negative slope in the Arrhenius plot) lowers with $T_A$, sign that the SBH decreases accordingly.

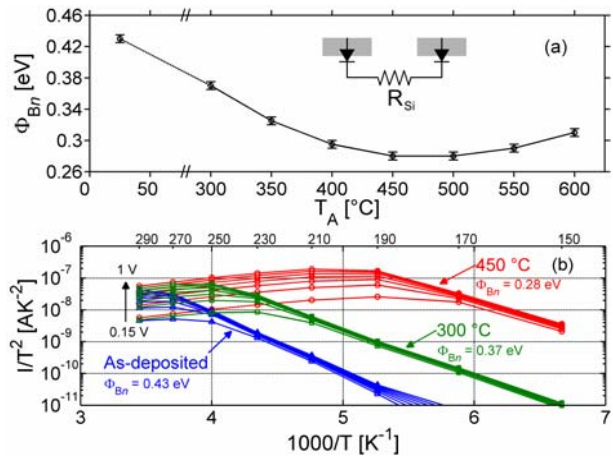

FIG. 1. (Color online) (a) SBH of Er silicide contacts to *n*-Si as a function of $T_A$, for the as-deposited sample and for the samples annealed for 2 min between 300 and 600 °C by steps of 50 °C. (b) Evolution of the Arrhenius plot at the indicated annealing temperatures.

The formed phases are identified by $\theta$-$2\theta$ x-ray diffraction (XRD). Figure 2 presents the XRD spectra for the as-deposited, 300 °C, and 450 °C samples. It must be mentioned that the Ti cap is not stripped after annealing. For the as-deposited sample, the intense peak registered at $2\theta = 32°$ matches the (002) peak of hexagonal Er. At 300 °C, the Er(002) peak disappears and a dominant peak at $2\theta = 30.4°$ ($d = 2.96$ Å) is now recorded. This peak does not belong to the diffraction pattern of body-centered cubic (BCC) Er sesquioxide ($Er_2O_3$). A similar peak ($d = 2.95$ Å) was previously reported for incompletely oxidized Er films[13,14] with a lower oxygen content than $Er_2O_3$ (denoted $ErO_x$, with x < 1.5). At 350 °C, three peaks of similar intensities are detected. These peaks can all be related to Er compounds. The peak at $2\theta = 27.1°$ can be associated to the (100) peak of hexagonal $ErSi_{2-x}$. In addition to the peak of $ErO_x$, another one appears ($2\theta = 29°$), which is relevant to the (222) peak of BCC $Er_2O_3$. In consequence, we can infer that Er atoms in close contact to Si preferentially react with Si, while upper Er layers rather react with oxygen to form $ErO_x$ and $Er_2O_3$. Indeed, some residual oxygen present in the annealing atmosphere can diffuse through the Ti cap during the thermal treatment and react with Er. When $T_A$ is incremented to 400 °C, the $ErSi_{2-x}(100)$ peak becomes dominant. More precisely, the intensity ratio between the $ErSi_{2-x}(100)$ and BCC $Er_2O_3(222)$ peaks is drastically increased, showing that the silicidation is now preponderant. Moreover, the $ErO_x$ peak is replaced by a peak ($2\theta = 30.7°$) relevant to Er-Si-O compounds. Actually, for a higher $T_A$, Si, the main diffusing species during Er silicide formation[15], is more likely to diffuse toward the surface of the Er layer where it can combine with Er and oxygen. At 450 °C, the $ErSi_{2-x}(100)$ peak prevails, with small amounts of Er-Si-O and $Er_2O_3$.

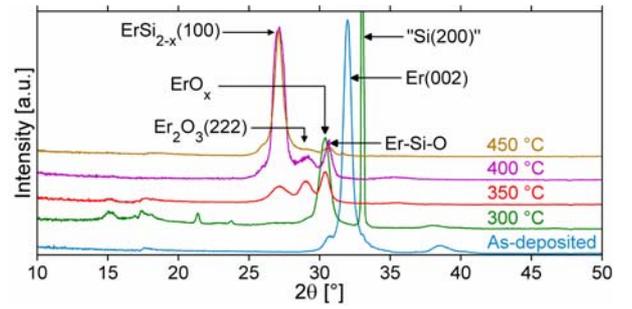

FIG. 2. (Color online) XRD spectra for the as-deposited sample and for the samples annealed for 2 min between 300 and 450 °C by steps of 50 °C, respectively.

With the aim of better understanding the SBH dependence on $T_A$, we inspect the metal-semiconductor (MS) interface. Figures 3(a)–3(c) exhibit high resolution transmission electron microscopy (HRTEM) micrographs of the samples considered in Fig. 1(b). In Fig. 3(a), a thin interfacial film (~3 nm) is discovered between Er and Si. In fact, Er and Si are known to be able to intermix even at room temperature to form an amorphous Er-Si (a-Er/Si) layer[16]. At 300 °C [Fig. 3(b)], we can see that the intermixed a-Er/Si layer continues to grow, reaching a thickness of approximately 12 nm. Simultaneously, crystalline $ErSi_{2-x}$ islands start to nucleate at the MS interface[17]. On the other hand, Er completely reacts with Si to form a crystalline $ErSi_{2-x}$ film at 450 °C [Fig. 3(c)], as already disclosed by the XRD measurements. From that analysis, it results that upon increasing $T_A$, the Er-Si alloy progressively evolves from an amorphous state of undetermined composition to a crystalline one with the $ErSi_{2-x}$ phase. Since the SBH accordingly decreases, we may believe that there is a correlation between the SBH lowering and the formation of crystalline $ErSi_{2-x}$.

Very few research groups have investigated the parameters that influence the SBH of RE silicides. One study in particular[18] provides some evidence for a correlation between the oxygen content in the silicide and the SBH. To further reinforce our previous assertion, the possible oxygen contamination of the MS interface must be examined. To do so, changes of chemical composition are explored by x-ray photoelectron spectroscopy (XPS) depth profiles performed with a monochromatized Al x-ray source and an $Ar^+$ sputter gun with an ion energy of 1 keV and a beam raster of $5\times5$ mm$^2$. Since the occurrence of interfacial oxygen has already been found to be under the XPS detection limit for $T_A = 450$ °C, we focus on the as-deposited and 300 °C samples. They are both introduced in the analysis chamber immediately after the evaporation to avoid oxidation in ambient air. Core level spectra are recorded for erbium (Er 4$d$), oxygen (O 1$s$), silicon (Si 2$s$), and titanium (Ti 2$p$). Figures 4(a) and 4(b) display the corresponding Er 4$d$, O 1$s$, and Si 2$s$ intensity profiles, whereas Ti 2$p$ is omitted for clarity. Figure 4(c) illustrates the evolution of the Er 4$d$ and Si 2$s$ BEs versus the sputter time (ST).

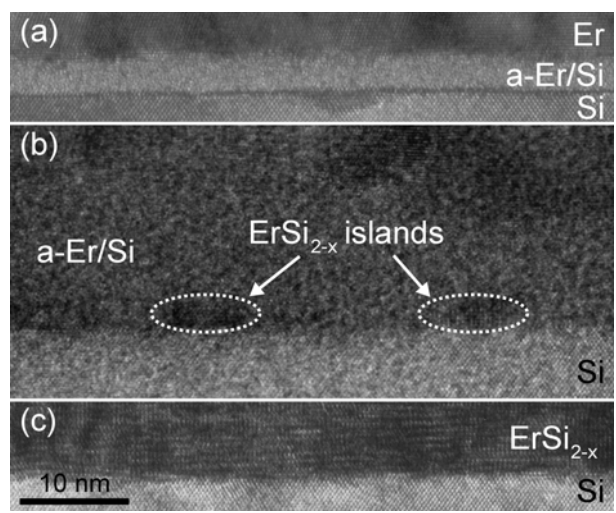

FIG. 3. HRTEM micrographs of the MS interface for the (a) as-deposited, (b) 300 °C, and (c) 450 °C samples, respectively.

From Fig. 4(a), it is disclosed that only the top of the Er layer is contaminated by a small amount of oxygen. Starting from marker A (ST = 40 min), the typical Er 4$d$ binding energy (BE) is equal to 167.1 eV, relevant to metallic Er. Above marker A, no oxygen can be detected, within the experimental accuracy. Si appears at marker B (ST = 84 min), as indicated by a corresponding Si 2$s$ BE of 149.5 eV. Since this emission line is not relevant to elemental Si (Si 2$s$ BE = 150.5-150.6 eV), it clearly suggests that a small part of the Er film is alloyed to Si, what is supported by the HRTEM pictures. Such a large Si 2$s$ shift of ~1.1 eV is comparable to the Si 2$p$ shift noticed for an a-Er/Si mixture where Si atoms are more diluted into Er than in ErSi$_{2-x}$[19,20]. In addition, the Er 4$d$ BE is found out to slightly shift toward 167.3 eV at marker B. Since it coincides with the detection of Si, it might be a supplementary signature of the Er-Si alloying.

At 300 °C [Fig. 4(b)], we can see that oxygen penetrates deeply into the sample. Roughly, between markers A' (ST = 64 min) and B' (ST = 124 min), Er is oxidized (Er 4$d$ BE = 169.5 eV), corresponding to the ErO$_x$ layer revealed by the XRD data. The layer located between positions B' and C' (ST = 176 min) is essentially composed of a mixture of Er silicide and Er oxide, with a decreasing proportion of Er oxide. Oxygen disappears in marker C'. After that position, the Si 2$s$ BE progressively shifts toward the peak relevant to elemental Si, reached in D' (ST = 220 min) [see Fig. 4(c)], which position marks the interface with the Si substrate. Concomitantly, the typical Er 4$d$ emission line amounts to 167.4 eV and starts to differ beyond D'. The layer between C' and D' can be very likely assimilated to the a-Er/Si layer with crystalline ErSi$_{2-x}$ inclusions disclosed by HRTEM in Fig. 3(b). In any case, it is established that the respective MS interfaces of the as-deposited and 300 °C samples are both oxygen-free, within the experimental accuracy.

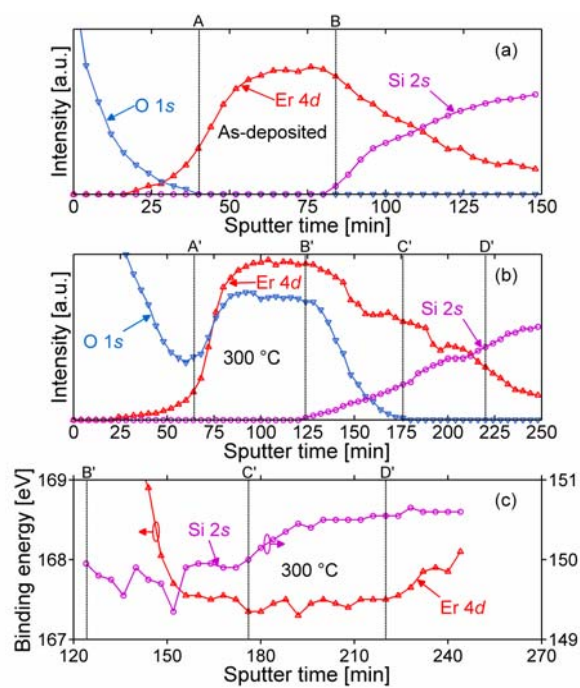

FIG. 4. (Color online) XPS depth intensity profiles for the (a) as-deposited and (b) 300 °C samples, respectively. (c) Er 4$d$ and Si 2$s$ BEs for the 300 °C sample.

In conclusion, we have shown that the SBH of Er silicide contacts to low doping $n$-Si exhibits a strong variation with $T_A$, dropping of ~0.15 eV after annealing at 450 °C. In the absence of interfacial contamination by oxygen, it proves out that this dependence originates from the progressive transformation of the Er silicide film under thermal annealing, leading to the formation of crystalline ErSi$_{2-x}$.

This work was supported by European Commission through the METAMOS project (METallic source/drain Architecture for Advanced MOS technology, IST-FP6-016677) and the NANOSIL network of excellence.